# Self-Tuning Control based on Modified Equivalent-Dynamic-Linearization Model

Feilong Zhang
State Key Laboratory of Robotics, Shenyang Institute of Automation, Chinese Academy of Sciences, Shenyang 110016, China

***Abstract*—The current model-free adaptive control (MFAC) method is designed by solving the optimization of a quadratic objective function. However, it does not take into account other system performance measures, such as zero-pole placement. Furthermore, this method fails to address the issues of time delay and disturbance in practical applications. To address these issues, i) we modified the EDLM by introducing time delay and disturbance, allowing a more accurate description of actual systems. Based on the modified EDLM, we proposed a class of self-tuning controllers to place the poles of the system and achieve the required system performance. Thereafter, we classified the proposed controller into four cases to enable easier applications. ii) The controller design and stability analysis of the system can be achieved by analyzing the function of the closed-loop poles. Furthermore, the issue of selecting the parameter $\lambda$ in the current MFAC based on quantity is resolved by analyzing zero-pole placement and the system's static error. In contrast, this can not be achieved using the invalid previous contraction mapping method; iii) the study of the proposed method focuses on the linear model to facilitate a better understanding of its working behavior. Finally, two examples are utilized to demonstrate the effectiveness of the proposed method and to point out the deficiencies in the current MFAC theory.**

## I. INTRODUCTION

MFAC was proposed in the 1990s and was not accepted by the control community until 2011. Over the past decade, a tremendous amount of work about MFAC has been published. The controller's design is based on an incremental form of the process model, known as the equivalent-dynamic-linearization model (EDLM), which includes a unit-time delay. Since time delays have not been taken into consideration, MFAC controllers designed based on EDLM may potentially fail in systems with delays [1]-[17]. Furthermore, most previous works on MFAC have been conducted without considering the influence of historical disturbances. In addition, [16] added one disturbance in full-form EDLM. Nevertheless, it neglected the impact of history disturbance. One motivation of this paper is to analyze MFAC for the discrete-time system subject to disturbance (white noise). To achieve this, a modified EDLM is proposed, incorporating time delay and disturbance to provide a more accurate description of the system. The proof is presented in the Appendix. Additionally, inspired by the design of a self-tuning controller, we propose a class of incremental forms of self-tuning control (ISTC) method based on the modified EDLM to achieve precise pole-zero placement of the system. Furthermore, Furthermore, we show that the current MFAC can be considered a special case of this ISTC method, as shown in Case 4. Both controllers satisfy the common stability analysis approach and the proven method for tracking error convergence. This is one of the most significant contributions and motivations in this research. Additionally, along with this, a different guideline for choosing the key parameter $\lambda$ is provided through quantitative analysis, which achieves a different conclusion from [1]-[10] that prove $\lambda$ should be sufficiently large to ensure the convergence of the tracking error through the contraction mapping method. Furthermore, Case 4 firstly demonstrates that static error of speed response will not be removed by MFAC when $\lambda\neq0$, which is confirmed by Example 2.1. Additionally, Example 2.2 shows that when the desired trajectory is $k^n$ ($0<n<\infty$), the tracking error converges to zero under $\lambda=0$. These indicate that the current analysis approaches are apparently unreasonable [1]-[17].

On the other hand, the STC in [18]-[22] is based on the linear combination of the input, output, and desired trajectory. It may require the introduction of an additional integrator to eliminate the influence of the constant disturbance and the underlying static error. In contrast, MFAC features a linear combination of incremental input, incremental output, and tracking error. This combination inherently includes one integrator. Moreover, ISTC shares the same structure as MFAC but also incorporates PID, designed in Case 2 and implemented in Example 1.2. This can help us enhance the behaviors of the system empirically.

The basic tool of MFAC is transforming the nonlinear system model to its EDLM using the definition of differentiability. This also implies that the process model EDLM represents a local linearization of a nonlinear system. Furthermore, we have demonstrated that the EDLM can be expressed by the ARMAX model in this paper. This accounts for the fact that the corresponding controller is essentially based on the linear model. Therefore, the study on the MFAC should begin with the linear system to more easily master its essence, regardless of whether the actual control system is linear or nonlinear. Adaptability to uncertainty or nonlinear systems can be achieved by combining this incremental form of the controller with an online estimation algorithm according to the certainty-equivalent principle.

The paper is organized as follows: Section II modifies the EDLM by introducing the disturbance and time delay. Then, the incremental form of STC is designed and classified into four cases to facilitate practical applications. Additionally, the relationship between the ARMAX model and EDLM is analyzed. The stability of the system is analyzed using the pole-zero placement method, which is also used for controller design. Section III presents several simulated examples to validate the authors' viewpoints and the effectiveness of the proposed controller. Section V gives the conclusion, and the Appendix contains the proof of the modified EDLM.

## II. Equivalent Dynamic Linearization Model with Unmeasured Stochastic Disturbances and Design of Incremental Form of Self-Tuning Control

### A. EDLM with Unmeasured Stochastic Disturbance

This section gives the EDLM with disturbance as a basic knowledge for the incremental self-tuning controller design. Its fundamental assumptions and theorem are given below.

We consider the following discrete-time SISO system:

$$\begin{aligned} y(k+1) = f(&y(k),\cdots,y(k-n_y),u(k-d+1),\cdots,u(k-d+1-n_u),\\ &w(k),\cdots,w(k-n_w)) + w(k+1) \end{aligned} \tag{1}$$

where $f(\cdots)\in R$ represents the nonlinear function, $n_y+1$, $n_u+1$, $n_w+1\in Z$ represent the unknown orders of the output $y(k)$, input $u(k)$ and the disturbance (or noise) $w(k)$ of the system at time $k$, respectively. $d$ represents the time delays between the input and the output.

Suppose system (1) conforms to below assumptions:

*Assumption 1*: The partial derivatives of $f(\cdots)$ with respect to all its variables are continuous.

*Assumption 2*: $w(k)$ is an uncorrelated random sequence of zero mean disturbance on the system.

*Theorem 1*: If the system (1) satisfies the above assumptions, there must exist a time-varying vector $\boldsymbol{\phi}_L(k)$ called PG vector; when $\Delta\boldsymbol{H}(k)\neq 0$, $L_y=n_y+1$, $L_u=n_u+1$ and $L_w=n_w+1$, the system (1) can be described into the following EDLM with disturbance:

$$\Delta y(k+1) = \boldsymbol{\phi}_L^T(k)\Delta\boldsymbol{H}(k) + \Delta w(k+1) \tag{2}$$

with $\|\phi_L(k)\|\le b$ for any time $k$, where

$$\boldsymbol{\phi}_L(k) = \begin{bmatrix} \boldsymbol{\phi}_{Ly}(k) \\ \boldsymbol{\phi}_{Lu}(k) \\ \boldsymbol{\phi}_{Lw}(k) \end{bmatrix} = [\phi_1(k),\cdots,\phi_{Ly}(k),\phi_{Ly+1}(k),\cdots,\phi_{Ly+Lu}(k), ,\phi_{Ly+Lu+1}(k),\cdots,\phi_{Ly+Lu+Lw}(k)]^T,$$

$$\Delta\boldsymbol{H}(k) = \begin{bmatrix} \Delta\boldsymbol{Y}_{Ly}(k) \\ \Delta\boldsymbol{U}_{Lu}(k-d+1) \\ \Delta\boldsymbol{W}_{Lu}(k) \end{bmatrix} = [\Delta y(k),\cdots,\Delta y(k-L_y+1), \Delta u(k-d+1),\cdots,\Delta u(k-d+1-L_u+1),\Delta w(k),\cdots,\Delta w(k-L_w+1)]^T$$

. *Proof*: Please refer to Appendix.

We define $\boldsymbol{\phi}_{Ly}(z^{-1}) = \phi_1(k)+\cdots+\phi_{Ly}(k)z^{-Ly+1}$, $\boldsymbol{\phi}_{Lu}(z^{-1}) = \phi_{Ly+1}(k)+\cdots+\phi_{Ly+Lu}(k)z^{-Lu+1}$, $\boldsymbol{\phi}_{Lw}(z^{-1}) = \phi_{Ly+Lu+1}(k)+\cdots+\phi_{Ly+Lu+Lw}(k)z^{-Lw+1}$ and $\Delta = 1-z^{-1}$.

*Assumption 3*: Suppose all roots of the polynomial $1+z^{-1}\boldsymbol{\phi}_{Lw}(z^{-1})=0$ are within the unit disk.

*Remark 1*: We will showcase the bidirectional transformation relationship between the ARMAX model and EDLM considering the presence of disturbances.

For the ARMAX model:

$$A(z^{-1})y(k+1) = z^{-d+1}B(z^{-1})u(k) + C(z^{-1})\zeta(k) \tag{3}$$

where $\zeta(k)$ is the bounded uncorrelated random sequence of zero mean disturbance with variance $\frac{\sigma^2}{2}$.

$A(z^{-1}) = 1+a_1z^{-1}+\cdots+a_{na}z^{-na}$, $B(z^{-1}) = b_0+\cdots+b_{nb}z^{-nb}$ and $C(z^{-1}) = 1+c_1z^{-1}+\cdots+c_{nc}z^{-nc}$ are polynomials in the unit delay operator $z^{-1}$; $n_a$, $n_b$, and $n_c$ are the orders of the system model (4). Letting (4)$-z^{-1}$(4), we have

$$\Delta y(k+1) = \alpha(z^{-1})\Delta y(k) + \beta(z^{-1})\Delta u(k) + \gamma(z^{-1})\Delta\zeta(k+1) \tag{4}$$

where

$$\alpha(z^{-1}) = -a_1z^{-1}-\cdots-a_{na}z^{-na}$$

$$\beta(z^{-1}) = b_0+\cdots+b_{nb}z^{-nb}$$

$$\gamma(z^{-1}) = 1+c_1z^{-1}+\cdots+c_{nc}z^{-nc}$$

Then let $\boldsymbol{\phi}_{Ly}(z^{-1}) = \alpha(z^{-1})$, $\boldsymbol{\phi}_{Lu}(z^{-1}) = \beta(z^{-1})$ and $(1+\boldsymbol{\phi}_{Lw}(z^{-1}))\Delta w(k+1) = \gamma(z^{-1})\Delta\zeta(k+1)$, we can obtain (3). This illustrates that the (4) can be expressed by (3).

Conversely, (2) can be rewritten as

$$(1-z^{-1}\boldsymbol{\phi}_{Ly}(z^{-1}))\Delta y(k+1) = z^{-d+1}\boldsymbol{\phi}_{Lu}(z^{-1})\Delta u(k) + (1+z^{-1}\boldsymbol{\phi}_{Lw}(z^{-1}))\Delta w(k+1) \tag{5}$$

We let $A(z^{-1}) = (1-z^{-1}\boldsymbol{\phi}_{Ly}(z^{-1}))\Delta$, $B(z^{-1}) = \boldsymbol{\phi}_{Lu}(z^{-1})\Delta$ and $C(z^{-1})\zeta(k) = (1+\boldsymbol{\phi}_{Lw}(z^{-1}))\Delta w(k+1)$, and then (5) can be described by (3).

### B. Design of Incremental Form of Self-Tuning Control

This section presents the ISTC design method and classifies the proposed controller into four cases to enable easier application. Most noticeably, we have analyzed the MFAC in Case 4 in a simple yet highly useful manner.

We can rewrite (2) into (6).

$$\begin{aligned} y(k+1) &= y(k) + \boldsymbol{\phi}_L^T(k)\Delta\boldsymbol{H}(k) + \Delta w(k+1) \\ &= y(k) + \boldsymbol{\phi}_{Ly}(z^{-1})\Delta y(k) + \boldsymbol{\phi}_{Lu}(z^{-1})\Delta u(k) \\ &\quad + \boldsymbol{\phi}_{Lw}(z^{-1})\Delta w(k) + \Delta w(k+1) \end{aligned} \tag{6}$$

A general incremental form of STC can be described by:

$$H(z^{-1})\Delta u(k) = E(z^{-1})\left[y^*(k+d) - y(k)\right] - G(z^{-1})\Delta y(k) \tag{7}$$

where, $H(z^{-1}) = h_0+h_1z^{-1}+\cdots+h_{nh}z^{-nh}$, $E(z^{-1}) = e_0+\cdots+e_{ne}z^{-ne}$ and $G(z^{-1}) = g_0+\cdots+g_{ng}z^{-ng}$ are polynomials. From (6) and (7), we can have the following equations.

$$\begin{aligned} y(k) &= \frac{z^{-d}\boldsymbol{\phi}_{Lu}(z^{-1})E(z^{-1})}{T}y^*(k+d) \\ &\quad + \frac{H(z^{-1})(1+z^{-1}\boldsymbol{\phi}_{Lw}(z^{-1}))}{T}\Delta w(k) \end{aligned} \tag{8}$$

$$u(k)=\frac{(1-\boldsymbol{\phi}_{Ly}(z^{-1}))E(z^{-1})}{T}y^{*}(k+d)-\frac{\left[E(z^{-1})+G(z^{-1})\Delta\right](1+z^{-1}\boldsymbol{\phi}_{Lw}(z^{-1}))}{T}w(k) \tag{9}$$

and the characteristic polynomial of the system.

$$T=H(z^{-1})(1-z^{-1}\boldsymbol{\phi}_{Ly}(z^{-1}))\Delta+z^{-d}\boldsymbol{\phi}_{Lu}(z^{-1})\left[E(z^{-1})+G(z^{-1})\Delta\right] \tag{10}$$

(8) and (9) present the closed-loop properties of the system in terms of the system output $y(k)$ and control input $u(k)$.

Given the desired closed-loop transfer function of the system is

$$G(z^{-1})=\frac{z^{-d}B_m(z^{-1})}{(1+z^{-1}\boldsymbol{\phi}_{Lw}(z^{-1}))(1-z^{-1})A_m(z^{-1})} \tag{11}$$

where $B_m(z^{-1})$ and $(1+z^{-1}\boldsymbol{\phi}_{Lw}(z^{-1}))(1-z^{-1})A_m(z^{-1})$ are the polynomials for desired zeros and poles of the closed-loop system, respectively. And the introduced polynomial $(1+z^{-1}\boldsymbol{\phi}_{Lw}(z^{-1}))(1-z^{-1})$ is the optimal observer on the basis of the optimal filtering theory [20]. To achieve the desired closed-loop transfer function of the system (11), $H(z^{-1})$, $E(z^{-1})$ and $G(z^{-1})$ should be such that the following two equations

$$\boldsymbol{\phi}_{Lu}(z^{-1})E(z^{-1})=B_m(z^{-1}) \tag{12}$$

$$\begin{aligned}(1+z^{-1}\boldsymbol{\phi}_{Lw}(z^{-1}))(1-z^{-1})A_m(z^{-1})&=z^{-d}\boldsymbol{\phi}_{Lu}(z^{-1})E(z^{-1})\\&+\left[H(z^{-1})(1-z^{-1}\boldsymbol{\phi}_{Ly}(z^{-1}))+z^{-d}\boldsymbol{\phi}_{Lu}(z^{-1})G(z^{-1})\right](1-z^{-1})\end{aligned} \tag{13}$$

where $\deg A_m(z^{-1})+L_w\le L_y+L_u+d-1$, $\deg G(z^{-1})=L_y$, $\deg H(z^{-1})=L_u+d-1$ and $\deg E(z^{-1})=\deg B_m(z^{-1})-L_u+1$.

Case 1: Ordinary controller design:

Without loss of generality for $L_y>L_u$ and $d=1$, we can rewrite (13) into (16) which is in the bottom of this page. Then $A_m(z^{-1})=a_0+a_1z^{-1}+\cdots+a_{na}z^{-na}$ and $B_m(z^{-1})=b_0+\cdots+b_{nb}z^{-nb}$ are the desired polynomials of poles and zeros of the closed-loop system, respectively. The stability of the system can be guaranteed when all the roots of $A_m(z^{-1})$ are located in the unit disk. Then $E(z^{-1})$ can be calculated from (12). $H(z^{-1})$ and $G(z^{-1})$ can be solved from (16). Then, the closed-loop poles will be shifted to the desired values specified by the polynomial $A_m(z^{-1})$ .

Case 2: Minimum phase system:

If the system is a minimum-phase system, meaning that all the roots of $\boldsymbol{\phi}_{Lu}(z^{-1})=0$ are in the unit circle, we can simplify the controller design method in Case 1. The system stability will be determined by

$$(1-z^{-1})\left[\left[1-z^{-1}\boldsymbol{\phi}_{Ly}(z^{-1})\right]H(z^{-1})+z^{-d}\boldsymbol{\phi}_{Lu}(z^{-1})G(z^{-1})\right]=T_1 \tag{14}$$

We can solve $H(z^{-1})$ and $G(z^{-1})$, which hold all roots of (14) in unit disk, to guarantee the system stability. Then we can tune the $E(z^{-1})=k_p(1-z^{-1})+k_i+k_d(1-2z^{-1}+z^{-2})$, which is in the PID form. To eliminate the influence of constant disturbance and steady error in the step response, we generally adopt $k_i\neq 0$. Then, the issue is converted into how to empirically tune the PID parameters in $E(z^{-1})$ to enhance the performance of a stable system.

Case 3: Deterministic system

If the disturbance is not considered, the EDLM with disturbance (2) is reduced to the following model [1], [9]:

$$\Delta y(k+1)=\boldsymbol{\phi}_L^T(k)\Delta\boldsymbol{H}(k) \tag{15}$$

$$\begin{bmatrix}
1 & 0 & \cdots & 0 & \phi_{Ly+1}(k) & 0 & \cdots & 0\\
\phi_1(k)-1 & 1 & \cdots & \vdots & \vdots & \phi_{Ly+1}(k) & \cdots & \vdots\\
 & \phi_1(k)-1 & \cdots & 0 & \phi_{Ly+Lu}(k)-\phi_{Ly+Lu-1}(k) & \vdots & \cdots & 0\\
\vdots & \vdots & \ddots & \vdots & -\phi_{Ly+Lu}(k) & \phi_{Ly+Lu}(k)-\phi_{Ly+Lu-1}(k) & \ddots & \\
\phi_{Ly}(k)-\phi_{Ly-1}(k) & & & \phi_1(k)-1 & 0 & -\phi_{Ly+Lu}(k) & & \phi_{Ly+1}(k)\\
-\phi_{Ly}(k) & \phi_{Ly}(k)-\phi_{Ly-1}(k) & & & 0 & 0 & & \vdots\\
0 & -\phi_{Ly}(k) & & & & 0 & \ddots & \\
\vdots & \vdots & \ddots & & \vdots & \vdots & & -\phi_{Ly+Lu}(k)\\
0 & 0 & & -\phi_{Ly}(k) & 0 & 0 & & 0
\end{bmatrix}
\begin{bmatrix}h_1(k)\\h_2(k)\\\vdots\\h_{nh}\\g_1(k)\\g_2(k)\\\vdots\\g_{ng}(k)\end{bmatrix}
=(1+z^{-1}\phi_{Lw}^T(z^{-1}))\begin{bmatrix}a_0\\a_1-a_0\\\vdots\\-a_{na}\\0\\\vdots\\0\end{bmatrix}-E(z^{-1})\begin{bmatrix}\boldsymbol{\phi}_{Lu}(k)\\0\\0\\\vdots\\0\end{bmatrix} \tag{16}$$

with $\|\boldsymbol{\phi}_L(k)\| \le b$ for any $k$, and

$$\boldsymbol{\phi}_L(k) = \begin{bmatrix} \boldsymbol{\phi}_{Ly}(k) \\ \boldsymbol{\phi}_{Lu}(k) \end{bmatrix} = [\phi_1(k), \cdots, \phi_{Ly}(k), \phi_{Ly+1}(k), \cdots, \phi_{Ly+Lu}(k)]$$

$$\Delta \boldsymbol{H}(k) = \begin{bmatrix} \Delta \boldsymbol{Y}_{Ly}(k) \\ \Delta \boldsymbol{U}_{Lu}(k-d+1) \end{bmatrix} = [\Delta y(k), \cdots, \Delta y(k-L_y+1), \Delta u(k-d+1), \cdots, \Delta u(k-d-L_u+2)]^T,$$

$$\boldsymbol{\phi}_{Ly}(z^{-1}) = \phi_1(k) + \cdots + \phi_{Ly}(k) z^{-Ly+1},$$

$$\boldsymbol{\phi}_{Lu}(z^{-1}) = \phi_{Ly+1}(k) + \cdots + \phi_{Ly+Lu}(k) z^{-Lu+1}.$$

We can have the following equations from (15) and (7).

$$y(k) = \frac{z^{-d} E(z^{-1}) \boldsymbol{\phi}_{Lu}(z^{-1})}{T_2(z^{-1})} y^*(k+d) \tag{17}$$

$$u(k) = \frac{E(z^{-1})\left[1 - z^{-d} \boldsymbol{\phi}_{Ly}(z^{-1})\right]}{T_2(z^{-1})} y^*(k+d) \tag{18}$$

where,

$$T_2(z^{-1}) = (1-z^{-1})\left[\left[1 - z^{-1}\boldsymbol{\phi}_{Ly}(z^{-1})\right] H(z^{-1}) + z^{-d}\boldsymbol{\phi}_{Lu}(z^{-1}) G(z^{-1})\right] + z^{-d}\boldsymbol{\phi}_{Lu}(z^{-1}) E(z^{-1}) \tag{19}$$

is the characteristic polynomial of the system.

The static error for the step response is

$$\lim_{k\to\infty} e(k) = \lim_{z\to 1} \frac{z-1}{z}\left(1 - \frac{E(z^{-1})\boldsymbol{\phi}_{Lu}(z^{-1})}{T_2(z^{-1})}\right)\frac{z}{z-1} = 0 \tag{20}$$

, indicating that the static error of the system for step response can be naturally eliminated using this class of method.

*Remark 2*: Whether we set $\Delta v(k) = (1 + z^{-1}\boldsymbol{\phi}_{Lw}(z^{-1}))\Delta w(k)$ for the simplification or not, the simulation experiment results show no difference if we design the controller according to Case 2. The EDLM is described by

$$y(k+1) = y(k) + \boldsymbol{\phi}_L^T(k)\Delta \boldsymbol{H}(k) + \Delta v(k+1) \tag{21}$$

where $v(k)$ in (21) can denote the external disturbance acting on the system output. From (7) and (21), we have

$$y(k) = \frac{z^{-d}\boldsymbol{\phi}_{Lu}(z^{-1})E(z^{-1})}{T_2} y^*(k+d) + \frac{H(z^{-1})}{T_2}\Delta v(k) \tag{22}$$

$$u(k) = \frac{(1-\boldsymbol{\phi}_{Ly}(z^{-1}))E(z^{-1})}{T_2} y^*(k+d) - \frac{\left[E(z^{-1}) + G(z^{-1})\Delta\right]}{T_2} v(k) \tag{23}$$

The transfer function that relates the output to the disturbance $v(k)$ is

$$G_1(z^{-1}) = \frac{H(z^{-1})(1-z^{-1})}{T_2} \tag{24}$$

Based on (24), it is evident that the system's static error, which arises from the constant disturbance of $v(k)$, can be effectively eliminated. Furthermore, by incorporating $m$ integrators into $H(z^{-1}) = (1-z^{-1})^m (h_0 + \cdots + h_{nh1} z^{-nh1})$, the impact of disturbance $v(k) = k^m$ will significantly diminish in theory.

Case 4: MFAC controller design

In this case, our objective is to present the approach for choosing the key parameter λ through quantitative analysis. This particular case is considered a special case of Case 3.

If $d=1$ and we choose $G(z^{-1}) = \phi_{Ly+1}(k)\boldsymbol{\phi}_{Ly}(z^{-1})$, $H(z^{-1}) = \lambda + \phi_{Ly+1}(k)\boldsymbol{\phi}_{Lu}(z^{-1})$ and $E(z^{-1}) = \phi_{Ly+1}(k)$, the controller will become the current MFAC. Then we can change the poles of the system by tuning the λ in (25).

$$T_3(z^{-1}) = (1-z^{-1})\lambda\left[1 - z^{-1}\boldsymbol{\phi}_{Ly}(z^{-1})\right] + \phi_{Ly+1}(k)\boldsymbol{\phi}_{Lu}(z^{-1}) \tag{25}$$

*Remark 3*: (a) The steady-state error (static error) in following the ramp input is

$$\begin{aligned} \lim_{k\to\infty} e(k) &= \lim_{z\to 1}\frac{z-1}{z}\left(1 - \frac{\phi_{Ly+1}(k)\boldsymbol{\phi}_{Lu}(z^{-1})}{T_3(z^{-1})}\right)\frac{T_s z}{(z-1)^2} \\ &= \lim_{z\to 1}\left(\frac{\lambda T_s\left[1-\boldsymbol{\phi}_{Ly}(z^{-1})\right]}{\lambda(1-z^{-1})\left[1-\boldsymbol{\phi}_{Ly}(z^{-1})\right] + \phi_{Ly+1}(k)\boldsymbol{\phi}_{Lu}(z^{-1})}\right) \end{aligned} \tag{26}$$

where, $T_s$ represents the sample time constant. We can conclude that the static error for the ramp response is proportional to $\lambda$. It is important to note that when $\lambda = 0$, we will have $\lim_{k\to\infty} e(k) = 0$. This conclusion contradicts the theorems in [1]-[10] that proved the convergence of the tracking error of the system controlled by MFAC is guaranteed under the condition that $\lambda$ is sufficient large. Furthermore, the convergence of the tracking error to zero in a system controlled by the current MFAC for the step response can be attributed to the inherent presence of an integrator in the controller, which is not determined by $\lambda$. (20) also supports this conclusion.

Besides, (26) also implies that the non-minimum phase system tracking error for speed response cannot converge to zero with current MFAC, since the system will be unstable if $\lambda = 0$. Furthermore, the static error in following the desired trajectory $k^n$ ($0 < n < \infty$) or any other trajectory $y^*(k) < \infty$ can be guaranteed to be zero theoretically by choosing $\lambda$=0 when the model is built precisely, and the stability is guaranteed, since

$$\begin{aligned} \lim_{k\to\infty} e(k) &= \lim_{\substack{z\to 1 \\ \lambda=0}}\frac{z-1}{z}\left(1 - \frac{\phi_{Ly+1}(k)\boldsymbol{\phi}_{Lu}(z^{-1})}{T_3(z^{-1})}\right)\frac{C(z)}{(z-1)^{n+1}} \\ &= \lim_{\substack{z\to 1 \\ \lambda=0}}\left(\frac{\lambda T_s\left[1-\boldsymbol{\phi}_{Ly}(z^{-1})\right]}{\lambda(1-z^{-1})\left[1-\boldsymbol{\phi}_{Ly}(z^{-1})\right] + \phi_{Ly+1}(k)\boldsymbol{\phi}_{Lu}(z^{-1})}\right)\frac{C(z)}{(z-1)^n} \\ &= 0 \end{aligned} \tag{27}$$

where, $Z(k^n) = \frac{C(z)}{(z-1)^{n+1}}$, $C(z)$ is the polynomial with the highest $n$ power and $Z(\bullet)$ denotes $z$-transformation.

(b) If the industrial settings require $\lambda \ne 0$ to improve the robustness against external disturbances or to avoid the denominator approaching zero, one option is to modify $H(z^{-1}) = \lambda + \phi_{Ly+1}(k)\boldsymbol{\phi}_{Lu}(z^{-1})$ of the current MFAC into $\lambda(1-z^{-1})^n + \phi_{Ly+1}(k)\boldsymbol{\phi}_{Lu}(z^{-1})$ to ensure the convergence of the system output to the desired trajectory $y^*(k) = k^n$ ($n<\infty$).

$$\lim_{k\to\infty} e(k) = \lim_{z\to 1}\frac{z-1}{z}\left(1-\frac{\phi_{Ly+1}(k)\boldsymbol{\phi}_{Lu}(z^{-1})}{T_4(z^{-1})}\right)\frac{C(z)}{\left(z-1\right)^{n+1}}$$
$$= \lim_{z\to 1}\left(\frac{\lambda(1-z^{-1})\left[1-\boldsymbol{\phi}_{Ly}(z^{-1})\right]C(z)}{\lambda(1-z^{-1})^{n+1}\left[1-\boldsymbol{\phi}_{Ly}(z^{-1})\right]+\phi_{Ly+1}(k)\boldsymbol{\phi}_{Lu}(z^{-1})}\right)$$
$$= 0 \tag{28}$$

where

$$T_4(z^{-1}) = \lambda(1-z^{-1})^{n+1}\left[1-z^{-1}\boldsymbol{\phi}_{Ly}(z^{-1})\right]+\boldsymbol{\phi}_{Lu}(z^{-1})\phi_{Ly+1}(k) \tag{29}$$

## III. Simulations

*Example* 1.1: This example aims to investigate the differences in ISTC implementation by comparing Case 2, which considers noise coefficients, with Case 3, which does not. Consider the following discrete-time SISO linear system:

$$y(k+1) = 1.5y(k) - 0.5y(k-1) + 0.1u(k-5) + 0.05u(k-6) + \xi(k) + 0.4\xi(k-1) \tag{30}$$

where $\xi(k)$ is an uncorrelated zero-mean random sequence with variance 0.01, and it is shown in Fig. 1. It is known that the exact coefficients of noise can be described by $\boldsymbol{\phi}_{Lw}(z^{-1}) = 1 + 0.4z^{-1}$. We assume that the process model in (31) accounts for noise coefficients as in Case 2, whereas (32) represents Case 3 where noise coefficients are not considered. The coefficients in both (32) and (33) are separately identified online using the least squares method, which are then utilized to determine the corresponding ISTC controllers.

$$y(k+1) = \hat{\phi}_1(k)y(k) - \hat{\phi}_2(k)y(k-1) + \hat{\phi}_3(k)u(k-5) + \hat{\phi}_4(k)u(k-6) + \xi(k) + \hat{\phi}_5(k)\xi(k-1) \tag{31}$$

$$y(k+1) = \hat{\phi}_1(k)y(k) - \hat{\phi}_2(k)y(k-1) + \hat{\phi}_3(k)u(k-5) + \hat{\phi}_4(k)u(k-6) \tag{32}$$

In this context, $\hat{*}$ represents the estimation of $*$, such as $\hat{\phi}$ represents the online estimated value of $\hat{\phi}$.

The desired output trajectory is

$$y^*(k+1) = 10\times(-1)^{round(k/100)}, 1\le k\le 400$$

The parameters and initial settings of the ISTC, based on the stochastic system (31) prescription in Case 2, and those of ISTC, based on the deterministic system (32) in Case 3, are shown in TABLE I. All the settings should be identical, except for the noise component. The estimation algorithm employs the least square method described in [20], with the tuning parameter set to $\boldsymbol{P}(0)=10^6\boldsymbol{I}$.

TABLE I Parameter Settings

| Parameter | Case 2 | Case 3 |
|---|---|---|
| Order | $L_y=2$, $L_u=2$, $L_w=1$ | $L_y=2$, $L_u=2$ |
| Initial value $\hat{\boldsymbol{\phi}}_L(1)$ | $[0.001, 0.001, 0.001, 0.001, 0.001]^T$ | $[0.001, 0.001, 0.001, 0.001]^T$ |
| $u(-3:3)$ | 0 | 0 |
| $y(0:2)$ | 0 | 0 |
| $E(z^{-1})$ | $0.5-0.3z^{-1}$ | $0.5-0.3z^{-1}$ |

Fig. 2 shows the tracking performance of the system under the control of both controllers. Fig. 3 displays the control input of both controllers. Fig. 4 and Fig. 5 depict the components of the PG estimation in Case 2 and Case 3, respectively.

From Fig. 2, we can see that the system controlled by incremental STC in Case 2 is slightly better than in Case 3 for less tracking error only within the time interval [0,100]. This improvement could be attributed to the estimated coefficient of noise item not immediately converging to the actual value, which unexpectedly has a positive impact. During the time interval [100,400], however, both systems demonstrate comparable control effectiveness.

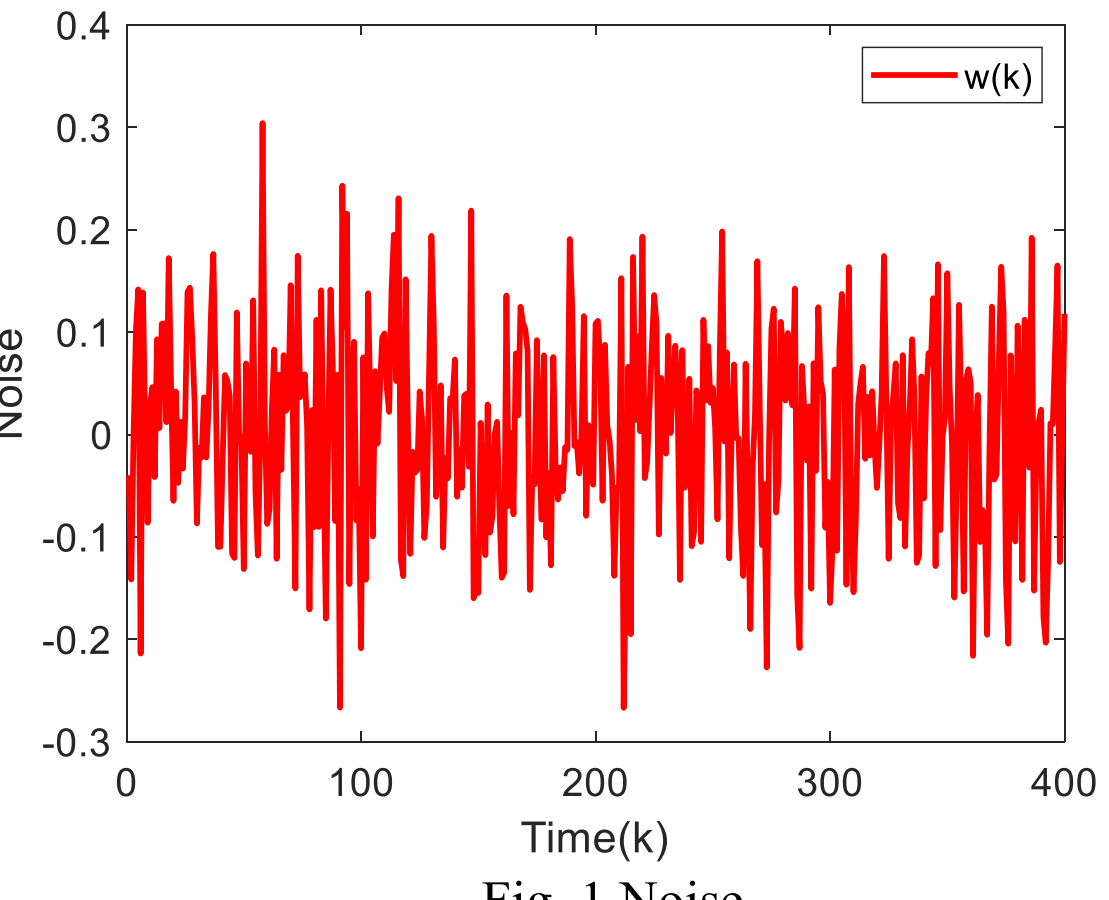

Fig. 1 Noise

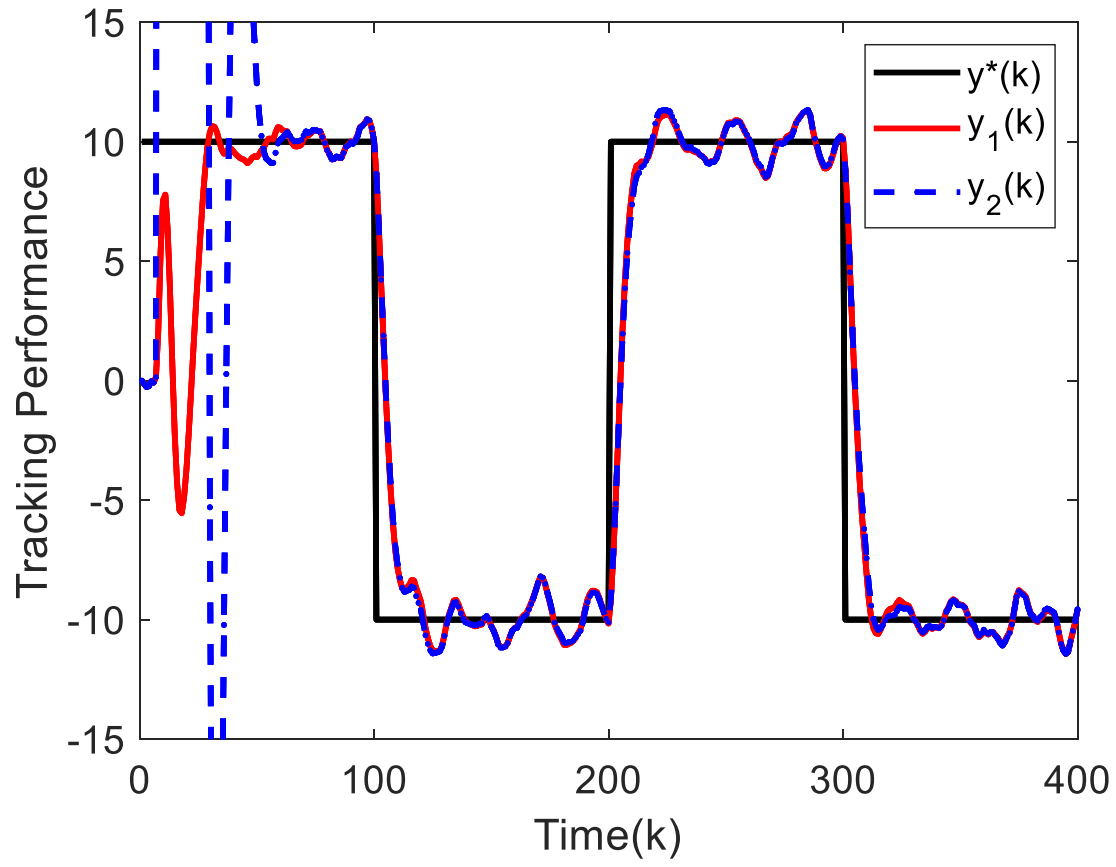

Fig. 2 Tracking performance

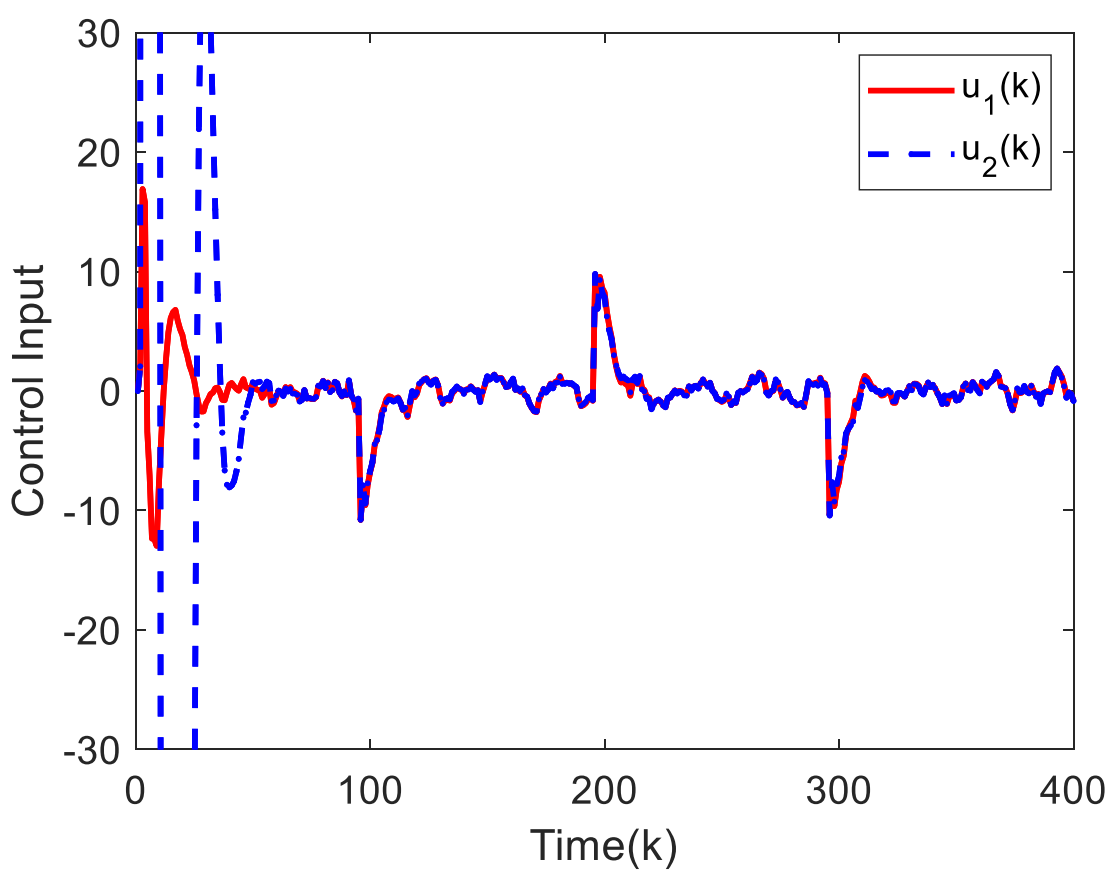

Fig. 3 Control input

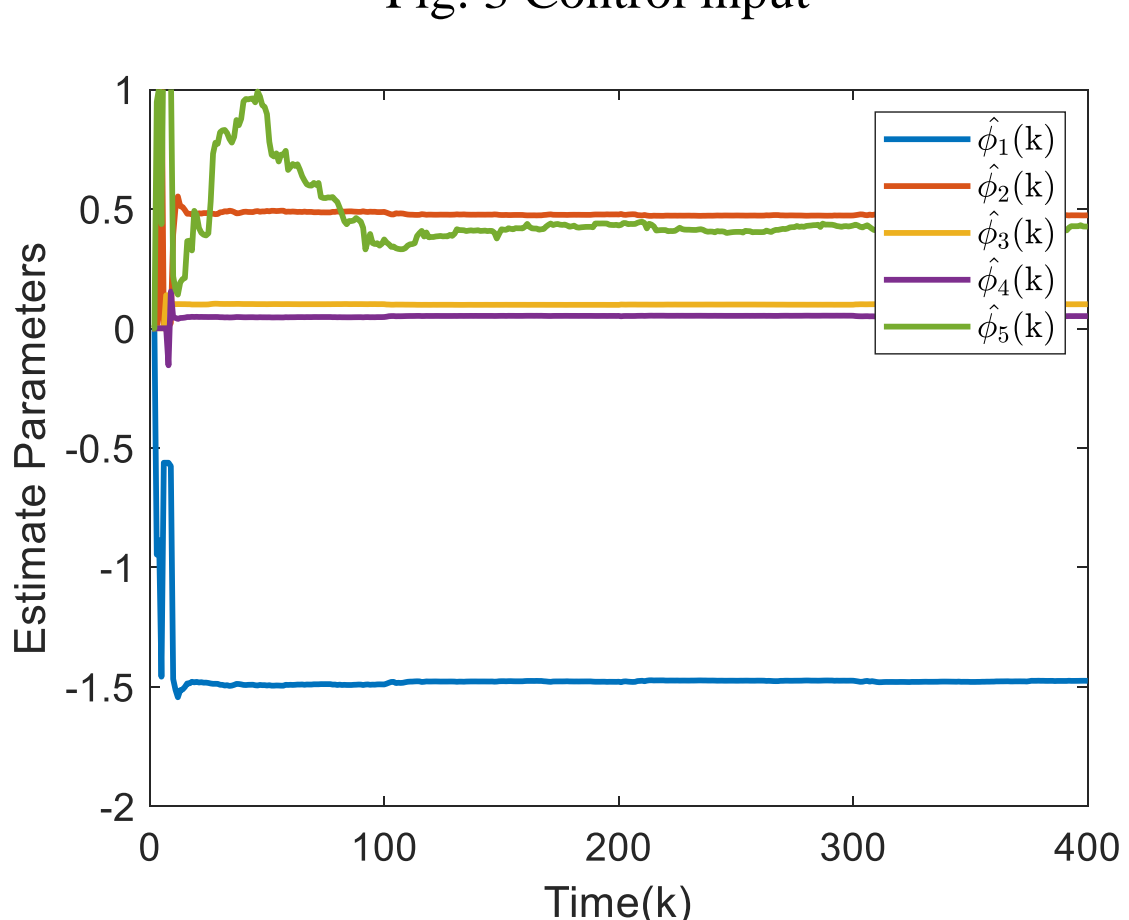

Fig. 4 Estimated value of PG considering noise

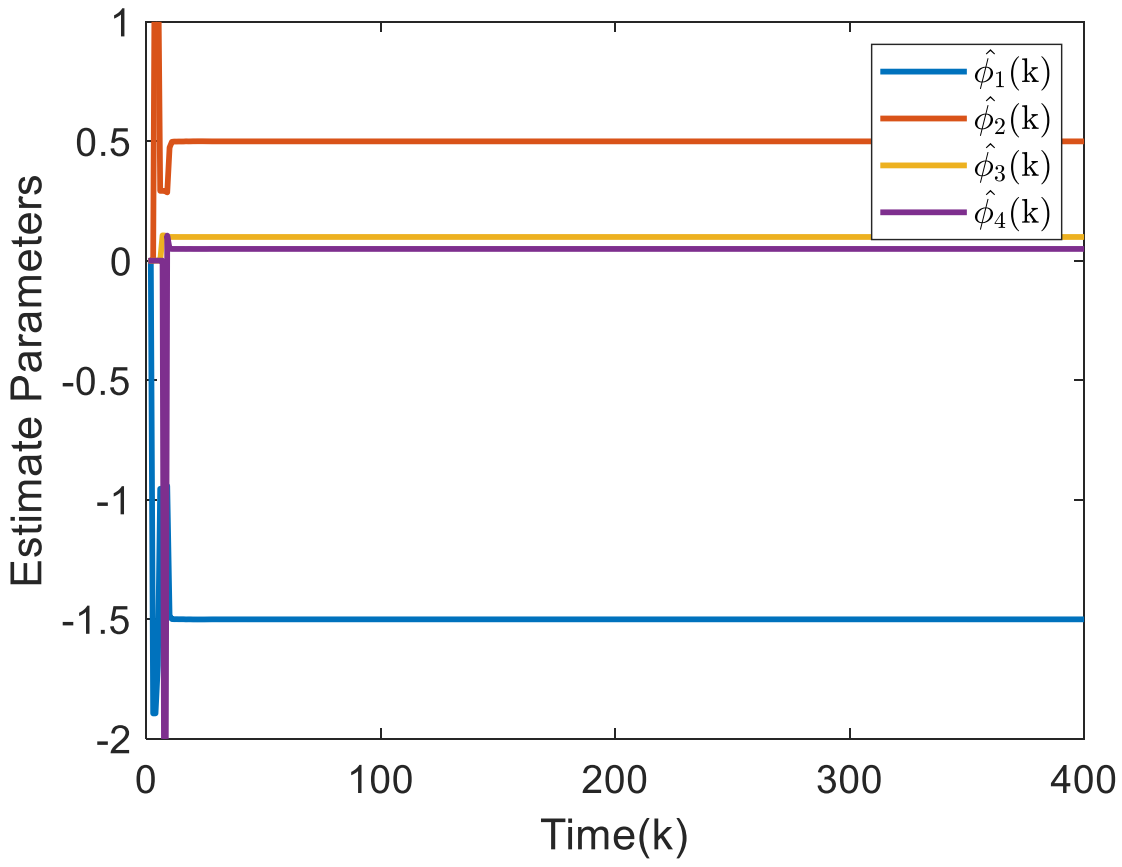

Fig. 5 Estimated value of PG not considering noise

*Example* 1.2: According to Case 2, this example aims to demonstrate the empirical methods used to tune the PID parameters in $E(z^{-1}) = k_p(1-z^{-1}) + k_i + k_d(1-2z^{-1}+z^{-2})$, to enhance the performance or change the system's behavior. This process is typically carried out after ensuring system stability, as indicated by (14).

The model is applied with (30) by removing noise, i.e., $\xi(k)=0$, to clearly observe the corresponding effects when different sets of PID are chosen in $E(z^{-1})$. Fig. 6 shows the tracking performance of the system controlled by the ISTC. The system outputs are: i) $y_s$: $E(z^{-1}) = 0.5-0.35z^{-1}$ ($k_P = 0.35$, $k_I = 0.15$); ii) $y_m$: $E(z^{-1}) = 0.5-0.3z^{-1}$ ($k_P = 0.3$, $k_I = 0.2$); ii) $y_l$: $E(z^{-1}) = 0.5-0.25z^{-1}$ ($k_P = 0.25$, $k_I = 0.25$);

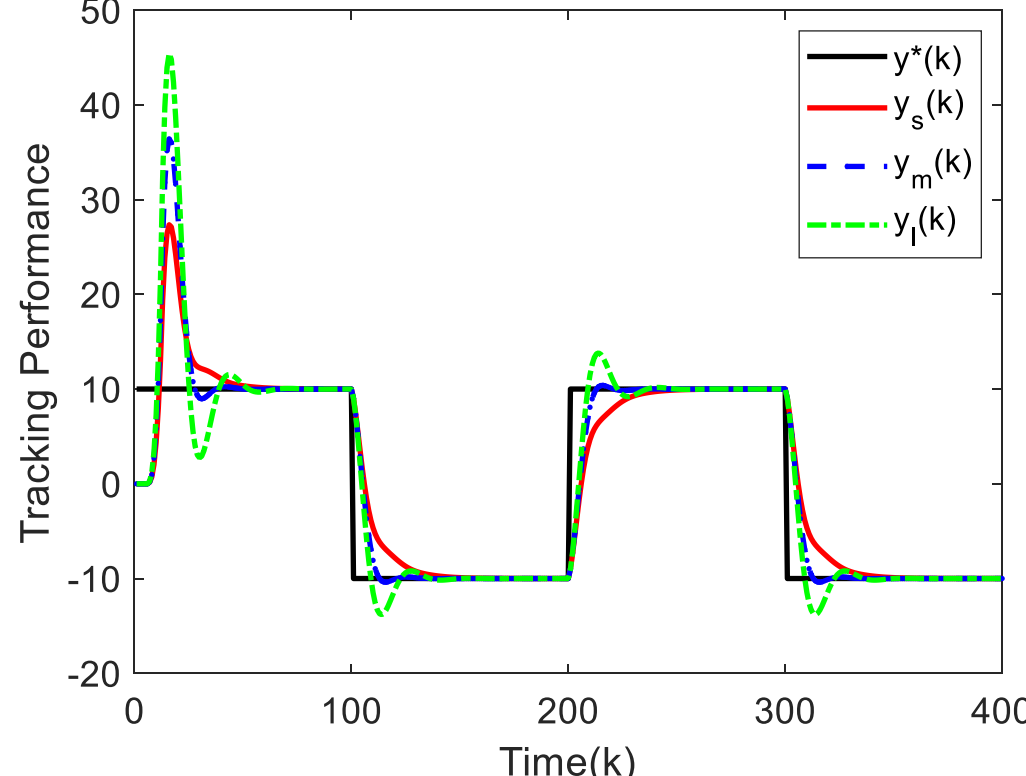

Fig. 6 Tracking performance

*Example* 2.1: Two simulations are conducted to validate the ideas about current MFAC in Case 4 in this example. The simulation firstly demonstrates that the static error in the speed response does not converge to zero with the current MFAC when $\lambda \neq 0$.

The first simulation investigates the method for eliminating the static error when $\lambda \neq 0$, as is required in Remark 3 (b). The discrete-time SISO linear controlled model is considered as

$$\Delta y(k+1) = -0.8\Delta y(k) - 0.5\Delta u(k) + 0.2\Delta u(k-1) \quad (33)$$

The desired trajectory is the unit-ramp signal. The controller parameters and initial settings for ISTC in Case 4 and the current MFAC are shown in TABLE II. They should be identical except for $H(z^{-1})$. The estimation algorithm uses the projection algorithm described in [1] with tuning parameters $\eta$ and $\mu$.

TABLE II Parameter Settings for Incremental STC and MFAC

| Parameter | Incremental STC | MFAC |
| --- | --- | --- |
| Order | $L_y = 1$, $L_u = 2$ | $L_y = 1$, $L_u = 2$ |
| $\eta$; $\mu$; $\lambda$ | 0.2; 1; 5 | 0.2; 1; 5 |
| Initial PG $\hat{\boldsymbol{\phi}}_L(1)$ | $[-0.1, -0.1, -0.1]^T$ | $[-0.1, -0.1, -0.1]^T$ |
| $u(0:6)$; $y(0:5)$ | 0; 0 | 0; 0 |
| $H(z^{-1})$ | $\lambda(1-z^{-1}) + \hat{\phi}_2(k)\hat{\boldsymbol{\phi}}_{Lu}(z^{-1})$ | $\lambda + \hat{\phi}_2(k)\hat{\boldsymbol{\phi}}_{Lu}(z^{-1})$ |

$G(z^{-1}) = \hat{\phi}_2(k)\hat{\boldsymbol{\phi}}_{Ly}(z^{-1})$, $E(z^{-1}) = \hat{\phi}_2(k)$.

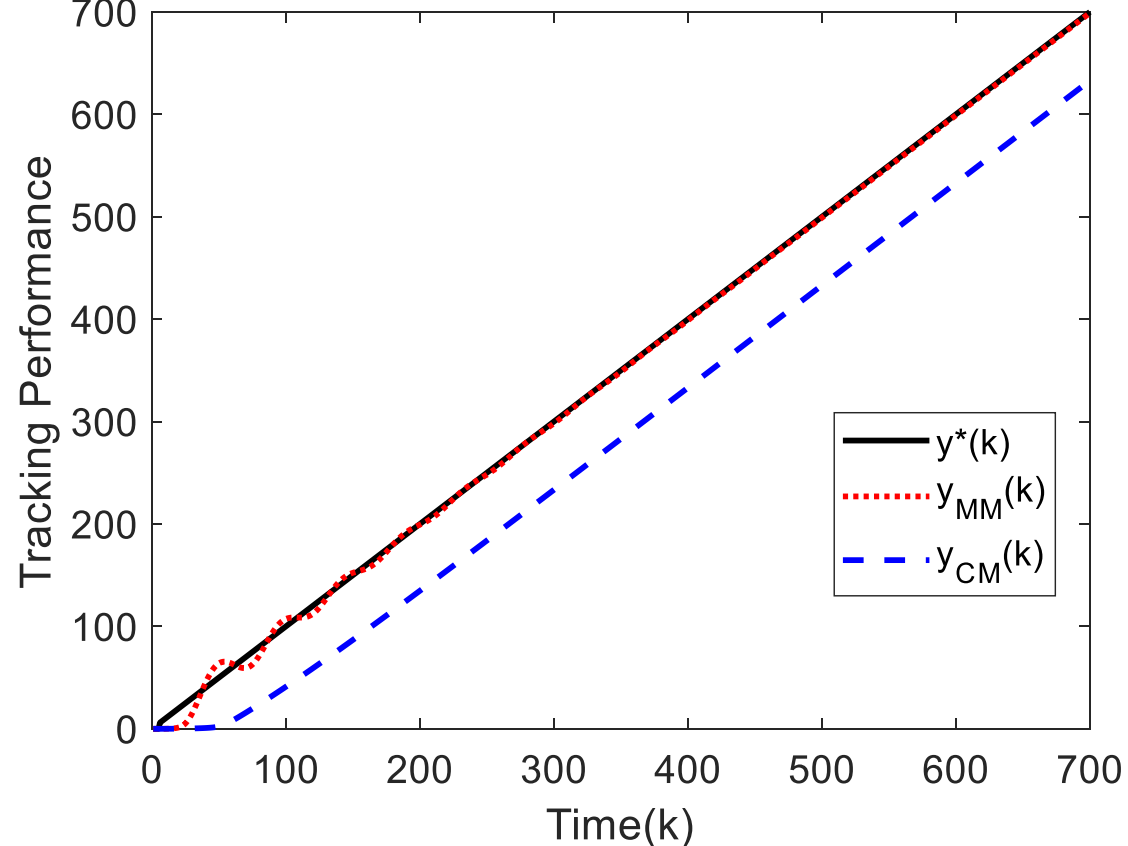

Fig. 7 Tracking performance

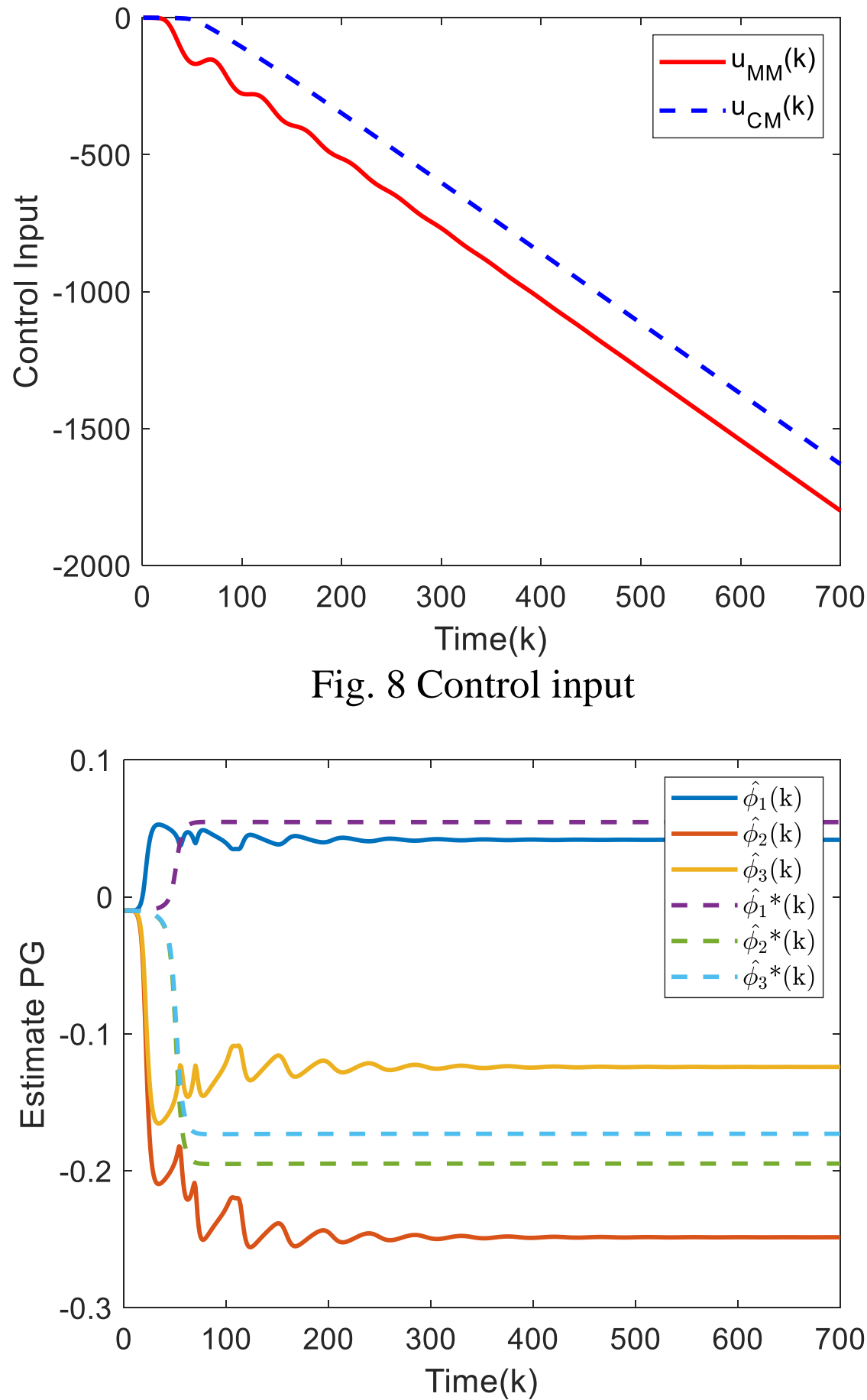

Fig. 8 Control input

Fig. 9 Estimated value of PG

Fig. 7 shows the tracking performance of the system under the control of both controllers. The outputs, $y_{MM}(k)$ and $y_{CM}(k)$, correspond to the system controlled by ISTC and the current MFAC, respectively. We can see that the static error of the system controlled by ISTC, designed based on the Remark 3 (b), has been successfully eliminated since the proposed method introduces an additional integrator in $H(z^{-1})$ compared to the current MFAC. Fig. 8 shows the control input of both. Fig. 9 presents the estimated PG vector components for the ISTC and MFAC, denoted as $\left[\hat{\phi}_1(k),\hat{\phi}_2(k),\hat{\phi}_3(k)\right]^T$ and $\left[\hat{\phi}_1^*(k),\hat{\phi}_2^*(k),\hat{\phi}_3^*(k)\right]^T$, respectively.

*Example* 2.2: We change the desired output trajectory with

$$y^*(k+1)=k^{10}, 1\le k\le 700 \tag{34}$$

to validate the conclusion in Remark 3 (a). Then, we apply the current MFAC with different values of $\lambda$. The tracking performance is shown in Fig. 10. The control inputs are shown in Fig. 11.

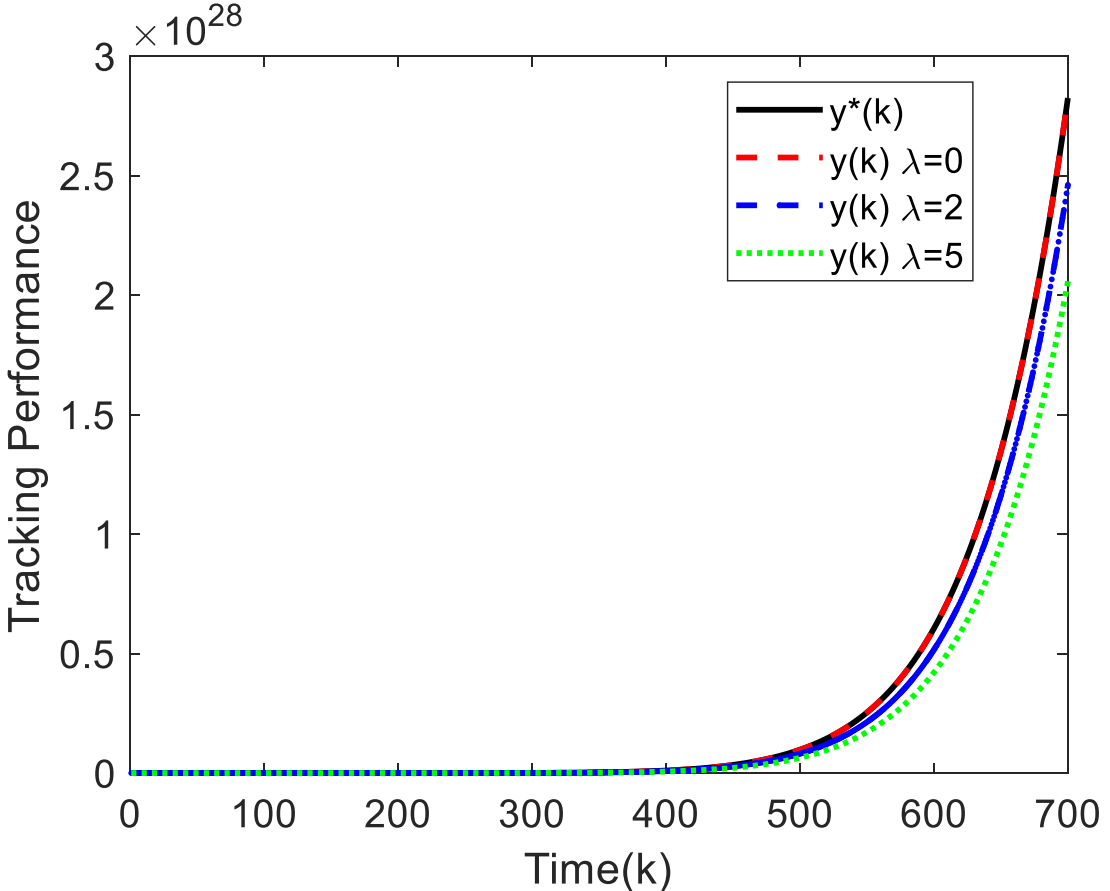

Fig. 10 Tracking performance

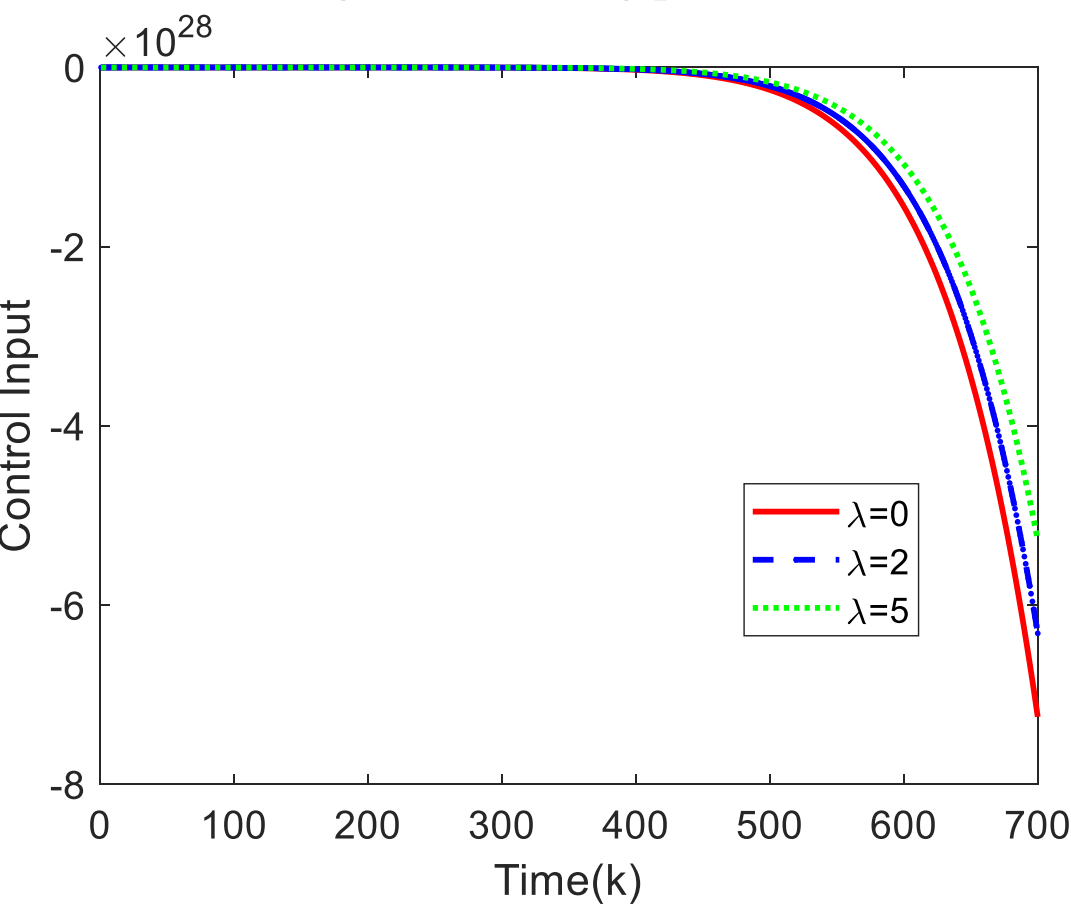

Fig. 11 Control input

From Fig. 10, it is evident that the static error will increase by raising the $\lambda$. Furthermore, we can conclude that the tracking error will not converge to zero for the speed response or accelerate response when $\lambda\neq 0$. Nevertheless, t it is worth noting that the tracking error for the desired trajectory $k^n$ ($0\le n<\infty$) may converge to 0 when $\lambda=0$. This discovery challenges the previously established conclusions in [1], [9].

## IV. Conclusion

**In this note, we present a family of incremental form of self-tuning method based on the modified EDLM, which is extended with the disturbance and time delay. The stability analysis of the system and controller design are completed by analyzing the function of closed-loop poles. We discuss several issues regarding the current MFAC. Several simulated examples are presented to validate the effectiveness of the proposed method and confirm the conclusions.**

## V. Appendix

### Proof of Theorem 2

*Proof*:

Case 1: $L_y=n_y+1$ and $L_u=n_u+1$, $L_w=n_w+1$.

According to the definition of differentiability, (1) can be rewritten as

$$\begin{aligned}\Delta y(k+1)=&\frac{\partial f(\boldsymbol{\varphi}(k-1))}{\partial y(k-1)}\Delta y(k)+\cdots+\frac{\partial f(\boldsymbol{\varphi}(k-1))}{\partial y(k-n_y-1)}\Delta y(k-n_y)\\ &\frac{\partial f(\boldsymbol{\varphi}(k-1))}{\partial u(k-d)}\Delta u(k-d+1)+\cdots\\ &+\frac{\partial f(\boldsymbol{\varphi}(k-1))}{\partial u(k-n_u-d)}\Delta u(k-n_u-d+1)\\ &\frac{\partial f(\boldsymbol{\varphi}(k-1))}{\partial w(k-1)}\Delta w(k)+\cdots+\frac{\partial f(\boldsymbol{\varphi}(k-1))}{\partial w(k-n_w-1)}\Delta w(k-n_w)\\ &+\gamma(k)+\Delta w(k+1)\end{aligned} \tag{35}$$

where,

$$\begin{aligned}\boldsymbol{\varphi}(k-1)=&[y(k),\cdots,y(k-n_y),u(k-d+1),\cdots,u(k-d+1-n_u),\\ &w(k),\cdots,w(k-n_w)]\end{aligned} \tag{36}$$

$$\begin{aligned}\gamma(k)=&\varepsilon_1(k)\Delta y(k)+\cdots+\varepsilon_{Ly}(k)\Delta y(k-n_y)\\ &+\varepsilon_{Ly+1}(k)\Delta u(k-d+1)+\cdots+\varepsilon_{Ly+Lu}(k)\Delta u(k-n_u-d+1)\\ &+\varepsilon_{Ly+Lu+1}(k)\Delta w(k)+\cdots+\varepsilon_{Ly+Lu+Lw}(k)\Delta w(k-n_w)\end{aligned} \tag{37}$$

and $\frac{\partial f(\boldsymbol{\varphi}(k-1))}{\partial y(k-i)}$, $\frac{\partial f(\boldsymbol{\varphi}(k-1))}{\partial u(k-d-j+1)}$, and $\frac{\partial f(\boldsymbol{\varphi}(k-1))}{\partial w(k-l)}$, denote the partial derivative values of concerning the *i*-th variable, the ($n_y+j$)-th variable and the ($n_y+n_u+l$)-th variable, respectively; $1\le i\le L_y$ $1\le j\le L_u$ and $1\le l\le L_w$ .And $\varepsilon_1(k),\cdots,\varepsilon_{Ly+Lu+Lw}(k)$ are functions that depend only on $\Delta y(k),\cdots,\Delta y(k-n_y),\Delta u(k),\cdots,\Delta u(k-n_u),\Delta w(k),\cdots,\Delta w(k-n_w)$ , with $(\varepsilon_1(k),\cdots,\varepsilon_{Ly+Lu+Lw}(k))\to(0,\cdots,0)$ when $(\Delta y(k),\cdots,\Delta y(k-n_y),\Delta u(k),\cdots,\Delta u(k-n_u),\Delta w(k),\cdots,\Delta w(k-n_w))$ $\to(0,\cdots,0)$ . This also implies that $(\varepsilon_1(k),\cdots,\varepsilon_{Ly+Lu+Lw}(k))$ can be regarded as $(0,\cdots,0)$ when the control period of the system is sufficiently small.

We let

$$\begin{aligned}\boldsymbol{\phi}_L(k)=[&\frac{\partial f(\boldsymbol{\varphi}(k-1))}{\partial y(k-1)}+\varepsilon_1(k),\cdots,\frac{\partial f(\boldsymbol{\varphi}(k-1))}{\partial y(k-n_y-1)}+\varepsilon_{Ly}(k),\\ &\frac{\partial f(\boldsymbol{\varphi}(k-1))}{\partial u(k-d)}+\varepsilon_{Ly+1}(k),\cdots,\frac{\partial f(\boldsymbol{\varphi}(k-1))}{\partial u(k-n_u-d)}+\varepsilon_{Ly+Lu}(k),\\ &\frac{\partial f(\boldsymbol{\varphi}(k-1))}{\partial w(k-1)}+\varepsilon_{Ly+Lu+1}(k),\cdots,\frac{\partial f(\boldsymbol{\varphi}(k-1))}{\partial w(k-n_w-1)}+\varepsilon_{Ly+Lu+Lw}(k)]^T\end{aligned} \tag{38}$$

and then (35) can be described as follows:

$$\Delta y(k+1)=\boldsymbol{\phi}_L^T(k)\Delta\boldsymbol{H}(k)+\Delta w(k+1) \tag{39}$$

The Case 2 ($1\le L_y\le n_y$, $1\le L_u\le n_u$ and $1\le L_w\le n_w$) and Case 4 ($L_y\ge n_y+1$ and $1\le L_u<n_u+1$, $1\le L_w\le n_w$; $0\le L_y<n_y+1$ and $L_u\ge n_u+1$, $1\le L_w\le n_w$;⋯) are similar to Case 1. Please refer to [23] for the analysis approach to save space.

We finished the proof of *Theorem 1*.